\documentclass[sigconf]{aamas} 
\usepackage{balance} % for balancing columns on the final page

\setcopyright{ifaamas}
\acmConference[AAMAS '24]{Proc.\@ of the 23rd International Conference
on Autonomous Agents and Multiagent Systems (AAMAS 2024)}{May 6 -- 10, 2024}
{Auckland, New Zealand}{N.~Alechina, V.~Dignum, M.~Dastani, J.S.~Sichman (eds.)}
\copyrightyear{2024}
\acmYear{2024}
\acmDOI{}
\acmPrice{}
\acmISBN{}

\acmSubmissionID{ea0117}

\title[Optimal Concessions with a Reservation Value]{A Negotiator’s Backup Plan: Optimal Concessions with a Reservation Value}

\author{Tamara C.P. Florijn}
\affiliation{
  \institution{Centrum Wiskunde \& Informatica}
  \city{Amsterdam}
  \country{The Netherlands}}
\affiliation{
  \institution{Utrecht University}
  \city{Utrecht}
  \country{The Netherlands}}
\email{Tamara.Florijn@cwi.nl}

\author{P{\i}nar Yolum}
\affiliation{
  \institution{Utrecht University}
  \city{Utrecht}
  \country{The Netherlands}}
\email{P.Yolum@uu.nl}

\author{Tim Baarslag}
\affiliation{
  \institution{Centrum Wiskunde \& Informatica}
  \city{Amsterdam}
  \country{The Netherlands}}
\affiliation{
  \institution{Utrecht University}
  \city{Utrecht}
  \country{The Netherlands}}
\email{T.Baarslag@uu.nl}
\begin{abstract}
Automated negotiation is a well-known mechanism for autonomous agents to reach agreements. To realize beneficial agreements quickly, it is key to employ a good bidding strategy. When a negotiating agent has a good back-up plan, i.e., a high reservation value, failing to reach an agreement is not necessarily disadvantageous. Thus, the agent can adopt a risk-seeking strategy, aiming for outcomes with a higher utilities.

Accordingly, this paper develops an optimal bidding strategy called MIA-RVelous for bilateral negotiations with private reservation values. The proposed greedy algorithm finds the optimal bid sequence given the agent's beliefs about the opponent in $O(n^2D)$ time, with $D$ the maximum number of rounds and $n$ the number of outcomes. The results obtained here can pave the way to realizing effective concurrent negotiations, given that concurrent negotiations can serve as a (probabilistic) backup plan.
\end{abstract}


\subtitle{Extended Abstract}

\newcommand{\BibTeX}{\rm B\kern-.05em{\sc i\kern-.025em b}\kern-.08em\TeX}
\usepackage{customstyle}

\makeatletter
\gdef\@copyrightpermission{
	\begin{minipage}{0.3\columnwidth}
		\href{https://creativecommons.org/licenses/by/4.0/}{\includegraphics[width=0.90\textwidth]{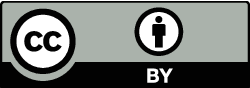}}
	\end{minipage}\hfill
	\begin{minipage}{0.7\columnwidth}
		\href{https://creativecommons.org/licenses/by/4.0/}{This work is licensed under a Creative Commons Attribution International 4.0 License.}
	\end{minipage}
	\vspace{5pt}
}
\makeatother

\begin{document}

%%% The following commands remove the headers in your paper. For final 
%%% papers, these will be inserted during the pagination process.

\pagestyle{fancy}
\fancyhead{}

%%% The next command prints the information defined in the preamble.

\maketitle 

%%%%%%%%%%%%%%%%%%%%%%%%%%%%%%%%%%%%%%%%%%%%%%%%%%%%%%%%%%%%%%%%%%%%%%%%

\section{Introduction}
Negotiation is often described as making concessions toward a mutually agreeable outcome \cite{young_1991_Negotiationanalysis}. Deciding on the right concessions is a challenge: being too stubborn can lead to non-agreements, while conceding too fast can result in bad agreements. An important factor in choosing the right negotiation strategy is that of a {\it backup plan}, which serves as a possible outcome that a negotiator can safely fall back on in case the negotiation ends in disagreement. Accordingly, this paper studies how the negotiator's backup plan can be used to design an optimal bidding strategy and how it influences a negotiator's concession behavior.

We consider the utility of this backup plan conceptually as a \textit{reservation value} \cite{mansour_2015_ApproachOnetoManyConcurrent,susskind_2011_WiggleroomRethinking, malhotra_2007_NegotiationGeniusHow, aydogan_2020_ANAC2018Repeateda,aydogan_2021_ANAC2017Repeated, curiel-cabral_2024_RiskAversionReservation}, interpreted as the value an agent receives if the negotiation fails \cite{williams_2014_OverviewResultsInsights, baarslag_2016_OptimalNonadaptiveConcession, mohammad_2021_Concurrentlocalnegotiations}. Various works consider how a reservation value can be used in a negotiation. A reservation value often functions as a lower bound of a bidder's utility target, signifying what can be obtained in other concurrent negotiation threads \cite{faratin_1998_Negotiationdecisionfunctions, raiffa_1982_artsciencenegotiation} or as a minimum acceptance threshold \cite{mirzayi_2022_opponentadaptivestrategyincrease}. Alternatively, reservation values are used to define heuristics in a multi-threaded negotiation context \cite{cuihongli_2005_DynamicOutsideOptions,williams_2012_NegotiatingConcurrentlyUnknown, yavuz_2020_TakingInventoryChanges}.  However, these works focus mostly on heuristic strategies and do not provide a theoretical proof of optimality.

In this work, we are interested in finding the optimal bidding curve given a private reservation value. The most closely related work is the Greedy Concession Algorithm (GCA)  \cite{baarslag_2015_OptimalNegotiationDecision}, which finds the theoretically optimal solution for time-sensitive domains with a static acceptance model, but without a reservation value. \citeauthor{mohammad_2023_Optimaltimebasedstrategy} \cite{mohammad_2023_Optimaltimebasedstrategy} has proven GCA can handle repeating offers. They have also introduced a faster GCA (QGCA) and extend to negotiations with a nonzero reservation value, but without direct proof for this setting.

Our main contribution is a bidding strategy called a Marginal Improvement Algorithm for Reservation Values (MIA-RVelous), which maximizes the marginal improvement of expected utility in a greedy way to find the optimal bid sequence given the reservation value. The algorithm finds an optimal sequence for static acceptance models with a time complexity of $O(n^2D)$. 

\section{Problem Setting}\label{sec:problem setting}
We focus on the bidding strategy (not the acceptance strategy or opponent model~\cite{Baarslag_2014_DecouplingNegotiatingAgents, baarslag_2016_Learningopponentautomated, razeghi_2020_Deepreinforcementlearning})  of a general agent for bilateral negotiation, where agents $A$ and $B$ take turns making bids, in accordance with the \textit{alternating offers protocol} \cite{rubinstein_1982_Perfectequilibriumbargaining}. If no agreement is reached before the maximum number of rounds, i.e. a deadline $D$, then the agent receives the reservation value $\rv$, which is considered private information and hence remains unknown to the opponent.

All the possible agreements together are called the \textit{negotiation domain} $\Omega = \{\w_1, \w_2, ...,\w_n\}$. Both agents have preferences over the domain $\Omega$; we adopt the perspective of agent $B$ and examine their bidding strategy. Agent $B$ associates with $\w_i \in \Omega$ a utility between $0$ and $1$, which we denote by $u(\w_i)$. $B$ has access to their own utility function, but $A$'s utility function is unknown to $B$. Therefore, $B$ uses an opponent model, with $a(\w_i)$ the acceptance probability of bid $\w_i$. 

\bexa
 Imagine Bo ($B$) would like to go for dinner together with their friend Ace ($A$), and they both have to agree on a restaurant before dinner time,  
 To come to the best dinner plan, Bo has to decide what bids to make and in what order. Suppose there are three options to consider: Italian, Sushi, and Fast food. In Table \ref{tab:examplesetting}, Bo's utility for each bid is shown, along with the corresponding (estimated) probability that Ace accepts the bid. Suppose Bo only has time for one dinner proposal out of the three options. Using Bo's belief regarding Ace's acceptance probabilities, Bo's first (and only) bid should maximize the expected utility, so the best choice is proposing the Fast food place, which yields an expected utility of $0.27$.
\eexa

\begin{table}[h]
\caption{Possible bids and their expected utilities for one bid.}
\begin{tabular}{p{2cm}p{1cm}p{2cm}p{2cm}}\toprule
\textbf{Bid}                     & \textbf{$B$'s utility} & \textbf{$A$'s acceptance probability} & \textbf{Expected \mbox{utility}}\\ \midrule
Italian & 0.5                & 0.4    & 0.20                           \\
Sushi    & 0.9                & 0.2   & 0.18                            \\
Fast food           & 0.3                & 0.9    &$0.27$
\\\bottomrule
\end{tabular}

\label{tab:examplesetting}
\end{table}

The example above illustrates one bid only; if we consider multiple bids strategically, the agent (Bo) can plan multiple bids ahead and decide on the whole bid sequence $\pi = (x_1,x_2, ...,x_k)$, with $x_1,...,x_k \in \Omega$ before the negotiation start given a static acceptance model. Note that early termination is irrational without bidding costs. $B$'s expected utility of making bids, $EU_\rv$ of $\pi$, is the expected utility of the first bid, or, if it is rejected, the expected utility of the second bid, and so on, ending with the reservation value if no agreement is reached:

\begin{equation*} 
EU_\rv(\pi) = \sum_{i=1}^{k} u_{i}\cdot a_{i}\prod_{j=1}^{i-1}(1-a_{j}) + \rv\cdot \prod_{j=1}^{k}(1-a_{j}).
\end{equation*}

The overall aim is to select the optimal sequence of $D$ bids \mbox{$\pi^*\in \Omega^D$} such that the expected utility $EU_\rv$ is maximized:

\begin{equation} \label{eq:argmax}
    \pi^* = \argmax\limits_{\pi\in \Omega^{D}} EU_\rv(\pi).
\end{equation}

A naive strategy to find this optimal sequence $\pi^*$ would be to evaluate all sequences from $\Omega^{D}$ and pick the sequence with the highest expected utility. As the agreement space grows larger, this quickly becomes infeasible. Therefore, we develop a novel strategy called MIA-RVelous that finds $\pi^*$ in $O(n^2D)$ time.

\section{Concession Behavior}
The MIA-RVelous algorithm, a variation of \cite{baarslag_2015_OptimalNegotiationDecision}, is presented in Algorithm \ref{algorithm MIArvelous}. Intuitively, MIA-RVelous works by first adding the reservation value as a special bid, and then greedily adding the bid with the best marginal improvement to utility in expectation. 
\begin{theorem}
   \label{theorem optimality}
   \mbox{\textbf{MIA-RVelous}} selects the optimal sequence of bids $\pi^* = \argmax\limits_{\pi \in \Omega^{D}} EU_\rv(\pi)$ given deadline $D$.
\end{theorem}
We can prove the optimality as formulated in Theorem \ref{theorem optimality}. To illustrate further, we extend our dinner example from Section~\ref{sec:problem setting}.
\begin{algorithm}[t]
    \caption{Marginal Improvement Algorithm with \rv\ \mbox{\textbf{(MIA-RVelous)}}}
    \label{algorithm MIArvelous}
        \begin{algorithmic}[1]
        \For{$k \in \{1,\dots,D\}$}
        \State $x_k \leftarrow \argmax\limits_{\omega\in \Omega \backslash \{x_1,\dots, x_{k-1}\}} EU_\rv(\sort(\{x_1, \dots, x_{k-1}, \omega\}))$
        \EndFor
        \State $\pi^* \leftarrow \sort(\{x_1, \dots, x_k\})$ \hfill \textit{//sorted by decreasing utility}
   \end{algorithmic}
\end{algorithm}

\bexa 
Bo and Ace negotiate about where to go for dinner. But this time, if the negotiation fails, Bo would eat a meal from the freezer, which yields a reservation value of $0.2$. Given the three options, what bids should Bo optimally propose? As shown before, the Fast food place should be chosen first without a backup plan. However, with a reservation value, it may not be immediately clear. If a bid's utility is lower than the reservation value, we can safely ignore the bid. Otherwise, we can use MIA-RVelous to determine what bid should be chosen with one available bid and a reservation value of 0.2, and apply Algorithm \ref{algorithm MIArvelous}, as shown below.

\begin{table}[h]
\caption{Calculations of the expected utility of Example 1.}
\begin{tabular}{rll}\toprule
\textbf{First bid} & \textbf{If rejected:} & \textbf{Expected utility} \\ \midrule
Italian  & \rv                   & $ 0.5 \cdot 0.4 + (1-0.4)\cdot 0.2 = 0.32. $                       \\
Sushi    & \rv                   & $0.9 \cdot 0.2 + (1-0.2)\cdot 0.2 = \fbox{0.34}.$                     \\
Fast food          & \rv                   & $0.3 \cdot 0.9 + (1-0.1)\cdot 0.2 = 0.29. $       \\              \bottomrule
\end{tabular}

\end{table}

Since the Sushi bid yields the highest expected utility, the first bid should be Sushi, which is more risky than Fast food meaning that it has a higher utility and a lower acceptance probability. This is an example of how the presence of a reservation value stimulates the agent to seek the riskier option. 
\eexa

\section{Discussion}\label{sec: discussion}
We can show that our proposed bidding strategy, MIA-RVelous, finds the optimal bid sequence for bilateral negotiations with a reservation value with a time complexity of $O(n^2D)$.

Our work leads to further research directions. In this paper, we consider a given reservation value, which we can extend to a scenario where the agent receives the reservation value with a certain probability. Probabilistic reservation values can serve their function in concurrent negotiations, as previously proposed~~\cite{cuihongli_2005_DynamicOutsideOptions}: an agent can fall back on ongoing concurrent negotiations after a negotiation thread failed, which results in probabilistic backup plans. Therefore, an important future direction is to extend our work on the optimal concession curve with a reservation value to one-to-many and many-to-many negotiations. The results obtained here can pave the way to realizing effective concurrent negotiations.

\begin{acks}
This publication is part of the Vidi project COMBINE (VI.Vidi.203.044) which is (partly) financed by the Dutch Research Council (NWO). 
\end{acks}

%%%%%%%%%%%%%%%%%%%%%%%%%%%%%%%%%%%%%%%%%%%%%%%%%%%%%%%%%%%%%%%%%%%%%%%%

%%% The next two lines define, first, the bibliography style to be 
%%% applied, and, second, the bibliography file to be used.
\balance
\bibliographystyle{ACM-Reference-Format} 
\bibliography{AAMAS-2024-Formatting-Instructions-CCBY_final/References}

%%%%%%%%%%%%%%%%%%%%%%%%%%%%%%%%%%%%%%%%%%%%%%%%%%%%%%%%%%%%%%%%%%%%%%%%

\end{document}